\documentclass[aps,
	       prd,
	       nofootinbib,
	       preprintnumbers,
	       nobibnotes,
	       twocolumn]{revtex4-1} 

\usepackage{amssymb}
\usepackage{amsmath}
\usepackage{graphicx,subfigure}
\usepackage{longtable}
\usepackage{verbatim}
\usepackage{amsfonts}
\usepackage{hyperref}
\usepackage{color}

\newcommand{\be}{\begin{equation}}
\newcommand{\ee}{\end{equation}}
\newcommand{\bea}{\begin{eqnarray}}
\newcommand{\eea}{\end{eqnarray}}
\newcommand{\beq}{\begin{equation}}
\newcommand{\eeq}{\end{equation}}
\def\beqa{\begin{eqnarray}}
  \def\eeqa{\end{eqnarray}}
\newcommand{\bv}{\left(\begin{array}{c}}
\newcommand{\ev}{\end{array}\right)}

\def\lsim{\mathrel{\rlap{\lower4pt\hbox{\hskip1pt$\sim$}}
    \raise1pt\hbox{$<$}}}	  
\def\gsim{\mathrel{\rlap{\lower4pt\hbox{\hskip1pt$\sim$}}
    \raise1pt\hbox{$>$}}}	  

\begin{document}


\title{The $V_{cb}$ puzzle: an update}
\author{Paolo Gambino}
\email{gambino@to.infn.it}
\author{Martin Jung}
\email{jung@to.infn.it}
\affiliation{
Dipartimento di Fisica, Universit\`a di Torino \& INFN, Sezione di Torino, I-10125 Torino, Italy}
\author{Stefan Schacht}
\email{ss3843@cornell.edu}
\affiliation{Department of Physics, LEPP, Cornell University, Ithaca, NY  14853, USA}

\vspace*{1cm}

\begin{abstract}
We discuss the impact of the  recent untagged analysis of ${B}^0\rightarrow D^{*}l\bar{\nu}_l$ decays by the Belle Collaboration on the  extraction of the CKM element 
$|V_{cb}|$ and provide updated SM predictions for the $b\to c\tau\nu$ observables $R(D^*)$, $P_\tau$, and $F_L^{D^*}$. The value of  $|V_{cb}|$ that we find is about $2\sigma$
from the one  from inclusive semileptonic $B$ decays, and is
very sensitive to the slope of the form factor at zero recoil which should soon become available from lattice calculations. 
\end{abstract}

\maketitle

\section{Introduction}

The  values of  the Cabibbo-Kobayashi-Maskawa matrix element $|V_{cb}|$ determined from 
inclusive and exclusive semileptonic $B$ decays have differed at the level of about $3\sigma$ for quite some time.
The situation around 2016 is well summarized by the latest HFLAV world average \cite{Amhis:2016xyh},
 \be 
|V_{cb}|= (39.05 \pm0.75)\ 10^{-3}\quad ({\rm HFLAV, }\, B\to D^* \ell\nu)\,,\label{hflav}
\ee
which combines many experimental results by BaBar, Belle and previous experiments
on $B\to D^* \ell\nu$ with the only unquenched lattice calculation of the form factor at zero-recoil (where the $D^*$ is produced at rest in the $B$ rest frame)  
available at the time \cite{Bailey:2014tva}.  The $B\to D\ell\nu$ channel tended to be considerably less precise.
Eq.~(1) should be compared with the most recent 
inclusive determination \cite{Gambino:2016jkc},
\be 
|V_{cb}|= (42.00\pm 0.64)\ 10^{-3}\qquad (B\to X_c \ell\nu),\label{incl}
\ee
based on  
a fit to the moments of various kinematic distributions in the Heavy Quark Expansion.
This $3\sigma$ discrepancy has become known as the {\it $V_{cb}$ puzzle}.

The last few years have seen a  series of new theoretical and experimental results, each bringing a new piece to the puzzle. 
First,  progress in the lattice QCD calculations of the form factors of $B\to D\ell\nu$ 
\cite{Lattice:2015rga,Na:2015kha} at small recoil (high $q^2$) improved significantly the exclusive determination based on that channel: a combined fit \cite{Bigi:2016mdz} to experimental \cite{Glattauer:2015teq,Aubert:2009ac} and lattice data 
for the $q^2$ distribution 
gives\footnote{Consistent results have also been obtained in \cite{ Aoki:2019cca}.}
\be 
|V_{cb}|= (40.49\pm 0.97)\ 10^{-3}\qquad (B\to D \ell\nu)\,,
\ee
in good agreement with both (\ref{hflav}) and (\ref{incl}).

Lattice calculations for the $B\to D^*\ell\nu$ channel have not yet reached the same level of maturity, and there are published results only for the form factor at zero recoil, which implies an extrapolation 
of experimental data,  until recently performed by the experimental collaborations using the Caprini-Lellouch-Neubert (CLN) parametrization \cite{Caprini:1997mu}.
Two years ago, the Belle collaboration published  a  preliminary tagged analysis with results for the $q^2$ and angular distributions of $B\to D^*\ell\nu$  \cite{Abdesselam:2017kjf}.   These unfolded distributions
allowed, for the first time,  independent fits to the form factors and $|V_{cb}|$ using different parametrizations, with  the surprising result that $|V_{cb}|$ could vary by as much as 6\% between the CLN  and the Boyd-Grinstein-Lebed (BGL) \cite{Boyd:1997kz} parametrizations, with the latter lifting $|V_{cb}|$ to agreement with (\ref{incl}) \cite{Bigi:2017njr, Grinstein:2017nlq,Jaiswal:2017rve}.
The CLN and BGL parametrizations are both built on the  analytic properties of the form factors, which together with the operator product expansion applied to correlators of two hadronic $\bar c b$ currents allow us to constrain  them significantly. 

These constraints can be made stronger using HQET relations between $B^{(*)}\to D^{(*)} $ form factors, known at $O(1/m,\alpha_s)$, supplemented by QCD sum rules for the subleading Isgur-Wise functions, see \cite{Caprini:1997mu,Bernlochner:2017jka} and references therein.  
Ref.~\cite{Caprini:1997mu} (CLN) provides a simplified parametrization  which includes these {\it strong} constraints, but the uncertainties related to missing higher order contributions and to the simplified parametrization have never been fully addressed before 2017. In 
 \cite{Bigi:2017jbd} two of us implemented  the {\it strong} unitarity bounds  taking into account recent lattice calculations as well as conservative estimates of the theoretical uncertainties, and showed that they reduce the BGL vs CLN discrepancy in $|V_{cb}|$, but do not eliminate it.

As emphasized in \cite{Bigi:2017njr}, the large observed parametrization dependence could be specific to
the only data set available at that time for the $B \to D^*\ell\nu$ differential distributions.
What was certainly clear from the 2017 
analyses was the pivotal role of
precise information on the form factors in the small recoil region, and in particular their zero-recoil slope. 

Several new relevant lattice calculations have appeared since 2016, 
showing that the lattice community has recognized 
 the crucial  importance of the form factor determination for a resolution of the puzzle.
 First, the HPQCD collaboration has computed the $B\to D^*\ell\nu$ form factor at zero recoil   \cite{Harrison:2017fmw}, confirming the result by FNAL-MILC, although with less precision.   Their recent  results for $B_s\to D_s^*\ell\nu$ \cite{McLean:2019sds}, obtained using the same action for the $b$ quark as for the lighter ones and thereby circumventing a large source of systematic uncertainty, 
appear very promising. Preliminary results for the  $B\to D^*\ell\nu$ form factors at non-zero recoil have been presented by the JLQCD \cite{Kaneko:2018mcr} and FNAL/MILC  \cite{Aviles-Casco:2019vin} collaborations. JLQCD has shown that large deviations from  HQS in $R_1$
 (a form factor ratio, see below) seem to be excluded, while FNAL/MILC has shown blinded results for all the four $B\to D^*\ell\nu$ form factors, albeit without complete systematic uncertainties. In particular, the slope of the combination of form factors that enters the differential rate in $w$, ${\cal F}(w)$, appears to be determined with good accuracy at zero recoil. 
Still concerning form factor calculations, improved and updated results at $q^2\leq0$ with Light Cone Sum Rules (LCSR) have been presented in \cite{Gubernari:2018wyi}.

On the experimental side, a significant effort has been devoted to reanalyzing Babar and Belle data. The  Babar collaboration recently published a new tagged analysis of  $B\to D^*\ell\nu$
using the BGL parametrization \cite{Dey:2019bgc}, performing an unbinned four-dimensional fit. The reported value of $|V_{cb}|$  is $(38.4\pm 0.2\pm 0.6\pm 0.6)\times 10^{-3}$, but,
unfortunately, their data are not yet available for independent analyses.
     
     The purpose of this  note is to discuss the latest development, namely a new untagged analysis of $B\to D^*\ell\nu$ 
by the Belle collaboration \cite{Abdesselam:2018nnh}, and to provide a first  
assessment of the global resulting situation.

  \begin{figure*}[t]
 \begin{center}
  \includegraphics[width=10.5cm]{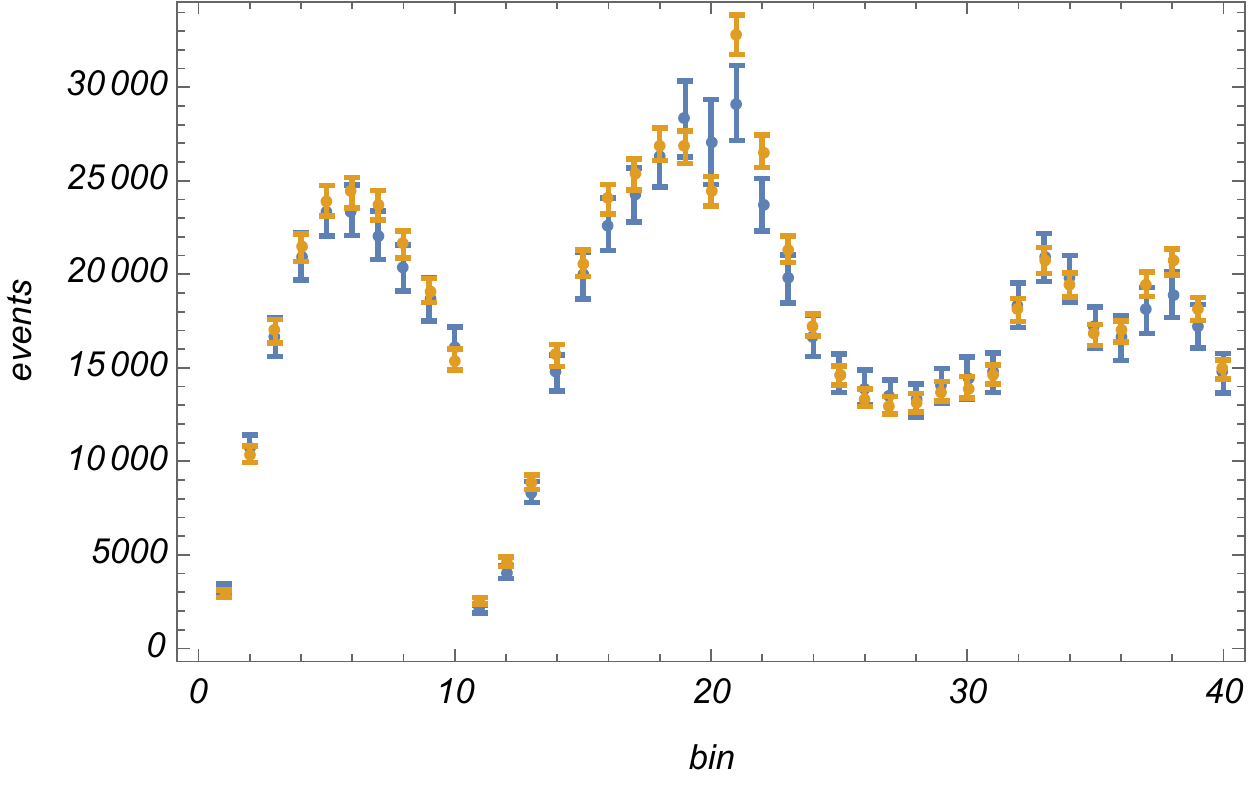} 
  \caption{Comparison between the Belle tagged \cite{Abdesselam:2017kjf} (blue points) and untagged \cite{Abdesselam:2018nnh} (orange points) datasets. The bins 1-10 refer to the $w$ spectrum, 11-20, 21-30 and 31-40  to the $\cos\theta_l$, $\cos\theta_v$ and $\chi$ distributions. }
\label{fig:compatibility}
\end{center}
\end{figure*}

The differential $B\to D^{(*)}\ell\nu$ data also provide bounds on new physics.
Indeed, the known differential distributions already place stringent bounds on all possible single-mediator models of new physics  \cite{Jung:2018lfu}, and none of these models can contribute to solve the $V_{cb}$ puzzle in a significant way \cite{Jung:2018lfu,Crivellin:2014zpa}, see also~\cite{Colangelo:2018cnj}. Even though it is unlikely to signal new physics, 
the $|V_{cb}|$ puzzle is still very  important because $i)$ it may be a signal that something is not yet understood
in either the inclusive or exclusive analysis,  with possible implications on the interpretation of
 $R(D^{(*)})$ as well, and $ii)$ it limits  the accuracy of $|V_{cb}|$ which affects  FCNC studies in an important way.

\section{Fitting the new data}
Like in Ref.~\cite{Abdesselam:2017kjf}, 
the authors of
Ref.~\cite{Abdesselam:2018nnh} provide one-dimensional distributions in the variables $w, \cos \theta_l, \cos\theta_v$, and $\chi$, with 10 bins for each variable. Unlike Ref.~\cite{Abdesselam:2017kjf}, however, they do not provide unfolded distributions, but give 
the efficiencies and response functions
 necessary to {\it fold} theoretical predictions in order to  get predictions for the yield in each bin. They also
perform binned fits to their data with both CLN and BGL parametrizations and find very similar results for $|V_{cb}|$, $(38.4\pm 0.2\pm 0.6\pm 0.6)\times 10^{-3}$ and 
$(38.3\pm 0.3\pm 0.7\pm 0.6)\times 10^{-3}$, respectively (here the errors are statistical, systematic, and due to the lattice form factor at zero recoil). 

These results are very different from those based on \cite{Abdesselam:2017kjf}: there is no sign of parametrization dependence in $|V_{cb}|$.  A first possibility is  then that the two datasets are incompatible. To investigate this point,
we take the unfolded distributions given in  \cite{Abdesselam:2017kjf}
and {\it fold} them in the experimental environment of Ref.~\cite{Abdesselam:2018nnh}, propagating the errors\footnote{The two analyses differ slightly in the endpoint for $w$, which leads to negligible differences in the angular bins but has a visible impact on bin 10.}. The bin by bin comparison of the yields for the sum of electrons and muons is shown in Fig.~\ref{fig:compatibility}. Despite visible  deviations in a couple of bins, the visual impression is of general compatibility. We will discuss  the issue in a  more  quantitative way when we will present fits including both datasets. Fig.~\ref{fig:compatibility} shows clearly that the 2018 data are considerably more precise than the 2017 ones. 

It is useful to recall at this point that in the BGL framework the four form factors relevant in $B\to D^*\ell\nu$, $f, g, F_1$ and $F_2$ (the latter entering only for $\ell=\tau$), are written
as power series in the variable $z=(\sqrt{w+1}-\sqrt{2})/(\sqrt{w+1}+\sqrt{2})$, where 
$w = (m_B^2+m_{D^*}^2-q^2)/(2 m_B m_{D^*})$, namely
\be 
f_i (z)= \frac{1}{P_{i}(z) \phi_{i}(z)} \sum_{k=0}^{n_i} a^{i}_k z^k. \label{BGLff}
\ee 
The {\it outer} functions $\phi_i(z)$ and the Blaschke factors $P_i(z)$ are given, for instance, in \cite{Bigi:2017njr,Bigi:2017jbd}.\footnote{
The $P_i(z)$ depend on the masses of the $B_c$ resonances below the lowest threshold. We use Table~III of \cite{Bigi:2017jbd} updating the masses of the second $0^-$ state  to 6.871GeV and of the second $1^-$ state to 6.910GeV due to their experimental discovery  \cite{Sirunyan:2019osb,Aaij:2019ldo,Eichten:2019gig}. This has a negligible impact on all of our results. The normalization of the outer function has been updated in \cite{Bigi:2017njr,Bigi:2017jbd}.} The $z$-expansion is truncated at order $n_i$, which may differ among different form factors.
The {\it weak} unitarity constraints on the parameters
$a_k^i$ are
\be
\sum_{k=0}^{n_g}  (a_k^g)^2<1,\quad \sum_{i=0}^{n_f} (a_k^f)^2+\sum_{k=0}^{n_{F_1}}(a_k^{{\cal F}_1})^2<1; 
\label {unitcon}
\ee
they ensure a rapid convergence of the $z$-expansion over the whole physical region, $0<z<0.056$.
The  Belle analysis \cite{Abdesselam:2018nnh} employs $n_f=1$, $n_g=0$, $n_{F_1}=2$, which we denote by BGL$^{(102)}$.\footnote{Notice that $a_0^{\mathcal{F}_1}$ is fixed by the value of~$a_0^f$, {\it cfr.}\ Eq.~(9) of \cite{Bigi:2017njr}. Note further that our notation BGL$^{(n_1,n_2,n_3)}$ corresponds to BGL$_{n_2+1,n_1+1,n_3}$ in Ref.~\cite{Bernlochner:2019ldg}.}

We have performed fits to the  2018 data of \cite{Abdesselam:2018nnh} using the information provided in that paper. A BGL$^{(102)}$ fit  without systematic errors  and using the same inputs  roughly reproduces the results reported in the paper, with $|V_{cb}|$ about 1\% higher, see Table~\ref{tab:fits}. The same applies to fits performed on the electron and muon data separately.  
Notice that, unless specified,  we always perform constrained fits subject to Eq.~(\ref{unitcon}).\footnote{This is different from adopting gaussian priors for each $a_k^{i}$, as was done {\it e.g.} in \cite{Lattice:2015rga}, which introduces a bias in the fit and affects the uncertainty in an uncontrolled way.} 
As expected, the inclusion of systematic errors and correlations has a considerable impact on the result of the fits, and it generally increases the central value of $|V_{cb}|$, see Table~\ref{tab:fits}.
Of course, a fit based on statistical uncertainties only, is likely to be biased,
because it gives too much weight to  
bins with larger systematic uncertainties, like those at small $w$ which depend on the soft pion reconstruction. On the other hand, the high degree of  correlation of the systematic errors 
for the angular bins suggests some caution. We will perform specific tests on the stability of the fit later on. For the moment, we observe that the complete covariance matrix does not show correlations exceeding 0.94 and that the correlations are generally slightly higher than in the 2017 analysis.

\begin{table}[t]
\begin{center}
\begin{tabular}{ccc|c|c}
\hline
\hline
data & fit & par	 & $\chi^2/\mathrm{dof}$ 	& $\vert V_{cb}\vert 10^3$	\\\hline 
2018 &stat &BGL$^{(102)} $& 53.0/35  &    $38.8\pm 0.6$\\
2018 &stat &CLN&  56.6/36  &    $39.2\pm 0.6$\\
2018 &naive &BGL$^{(102)} $& 32.6/35  &    $39.7\pm 0.9$\\
2018 &naive&CLN & 32.4/36  &    $39.7\pm 0.9$\\
2018 & &BGL$^{(102)} $& 32.5/35  &    $40.3\pm 0.9$\\
2018 & &CLN & 32.4/36  &    $40.3\pm 0.9$\\
2018 &stat &BGL$^{(222)}$ & 47.7/32 & $37.6^{+1.0}_{-0.9}$\\
2018& naive &BGL$^{(222)}$ & 31.2/32 & $38.6^{+1.5}_{-1.4}$\\
$2018$&  &BGL$^{(222)}$ & 31.2/32 & $39.1^{+1.5}_{-1.3}$\\ \hline
2017/18, slope& & BGL$^{(222)}$ &84.5/73 & $40.8 \pm 0.8$\\
2017/18, LCSR& & BGL$^{(222)}$ &80.5/75 & $39.5 \pm 1.0$\\
2017/18, LCSR, slope& & BGL$^{(222)}$ &88.0/76 & $40.8 \pm 0.8$\\
\end{tabular}
\caption{Results of various fits to the $B\to D^* \ell\nu$ data. The $|V_{cb}|$ error always includes the lattice uncertainty.  In the second column {\it stat} stands for "only statistical errors", {\it naive} stands for "systematic errors as fractions of the yield", in all other cases we 
take the D'Agostini effect into account (see text). All fits are 
with weak unitarity constraints.
\label{tab:fits} } 
\end{center}
\end{table}

An important point concerns the implementation of the systematic uncertainties, which Ref.~\cite{Abdesselam:2018nnh} provides as relative uncertainties. It is well-known that 
computing  the systematic uncertainty as a fraction of the yield in each bin can lead to a bias (the D'Agostini effect \cite{DAgostini:1993arp}), which  however can be avoided by expressing the systematic uncertainty 
as a fraction of the predicted yield. Indeed we observe a significant shift in $|V_{cb}|$ due to this effect, see Table~\ref{tab:fits}. There is a residual ambiguity depending on the 
form factors employed to predict the yields, but it is numerically very small, as the main effect is related to employing a prediction that is not subject to  fluctuations. In Table~I and in the following we will always compute the systematic errors from predictions based on the form factors obtained in a  fit where the systematic errors are a fraction of  the yields, unless explicitly stated.

In BGL fits the power of $z$ at which the series  in
(\ref{BGLff}) is truncated is potentially 
important for the extraction of $|V_{cb}|$.
For instance, the optimal choice of $(n_f, n_g,n_{F_1})$ in BGL fits has been recently discussed 
in \cite{Bernlochner:2019ldg}. 
We believe that, generally speaking,  the problem has a simple
solution: the optimal truncation of the $z$-expansion occurs when adding more terms does not change the result of the fit in any relevant way. Eqs.~\eqref{unitcon} together with $0\leq z\leq0.056$ guarantee the convergence of such a procedure. 
 Although this may imply adding (almost) redundant parameters subject to Eq.~(\ref{unitcon}), it is crucial for determining the uncertainty of $|V_{cb}|$ in a reliable way. 
We illustrate the point by comparing the BGL$^{(102)}$ fit with a BGL$^{(222)}$ fit,
having checked that nothing changes by adding even more parameters\footnote{In fact, we find that  BGL$^{(212)}$ leads to  results very similar to those of  BGL$^{(222)}$,
but to ease comparisons we stick to the choice made in \cite{Bigi:2017njr}.}. The total uncertainty increases from 
0.9 to 1.4 $10^{-3}$, which we think is  the correct uncertainty of $|V_{cb}|$ in  a BGL fit. 
The argument that a certain parameter can be dropped because 
the fit is unable to constrain it effectively is, in this particular case, ill-conceived. 
The BGL parametrization is not model-independent if one arbitrarily drops parameters. From now on we will limit ourselves to BGL$^{(222)}$ fits only.

Ref.~\cite{Bernlochner:2019ldg} also mentions the risk of overfitting.
Imposing at least {\it weak} unitarity, which is avoided in \cite{Bernlochner:2019ldg}, minimizes  this risk and is completely safe, because  the unitarity constraints (\ref{unitcon}) are  very far from being saturated by the $B\to D^*$ channel alone, see \cite{Bigi:2017jbd}. 

\begin{figure}[t]
	  \centering{
  \includegraphics[width=8.6cm]{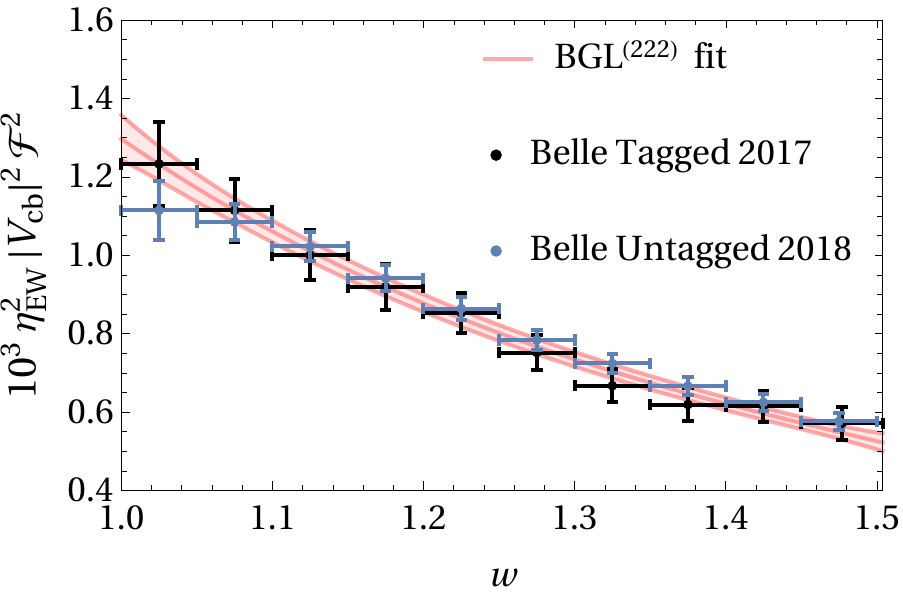} 
  \caption{Comparison of BGL$^{(222)}$ fit including weak unitarity constraints with the two Belle data sets.}
\label{fig:VcbFsqPlot}
}
\end{figure}

Let us now consider a fit to the combined 2017  \cite{Abdesselam:2017kjf} and 2018 \cite{Abdesselam:2018nnh} Belle datasets. Unlike the previous fits, where we were comparing directly with   \cite{Abdesselam:2018nnh}, we now employ the FLAG average for the form factor at zero-recoil, $h_{A_1}(1)=0.904(12)$ \cite{Aoki:2019cca}.
The complete results of this fit are given in Table~\ref{tab:fit-final}: they show a marked increase in the minimal $\chi^2/dof$, implying some tension between the 2017 and 2018 data. 
Nevertheless, the combined fit still has an excellent $p$-value of $\sim24\%$.
In Fig.~\ref{fig:VcbFsqPlot} we compare our fit result for $\eta^2_{\mathrm{EW}} \vert V_{cb}\vert^2 \vert\mathcal{F}(w)\vert^2$ with the two Belle data sets.
In order to show the data points of Ref.~\cite{Abdesselam:2018nnh} in the same plot with those of 
Ref.~\cite{Abdesselam:2017kjf}, we employ an effective bin-by-bin rescaling factor obtained by comparing yields and binned differential branching fraction in the case of our best fit.
We have performed a few checks on the stability of this fit: first, we have removed 
a few bins from the 2018 analysis, aiming at eliminating the strongest systematic correlations, and we did not observe any relevant change in  $|V_{cb} |$. If we  remove all
angular bins we get almost the same central value with larger uncertainty: $|V_{cb} |=(39.8\pm 1.5) 10^{-3}$. 
The $w$ bins are more important for the determination of $|V_{cb} |$, and the first two in particular; the fit prefers lower $|V_{cb} |$ only if we remove the first two $w$ bins of both 2017 and 2018 analyses, otherwise it is almost unchanged.

\begin{table}[t]
\begin{center} 
\begin{tabular}{c|c|c}
\hline
\hline
BGL$^{(222)}$ 		 & Data + lattice (weak)	& Data + lattice (strong) 	\\\hline 
$\chi^2/\mathrm{dof}$    & $80.1/72$ 		& $80.1/72$				\\\hline
$\vert V_{cb}\vert 10^3$      & $39.6\left(^{+1.1}_{-1.0}\right)$ 	& $39.6\left(^{+1.1}_{-1.0}\right)$ 		\\\hline
$a_0^f$	         	 & $0.01221(16)$			& $0.01221(16)$ 				\\ 
$a_1^f$	         	 & $0.006\left(^{+32}_{-45}\right)$     & $0.006\left(^{+20}_{-32}\right)$		\\
$a_2^f$	         	 & $ -0.2\left(^{+12}_{-8}\right)$      & $-0.2\left(^{+7}_{-3}\right)$			\\\hline
$a_1^{\mathcal{F}_1}$	 & $0.0042\left(^{+22}_{-22}\right)$ 	& $0.0042\left(^{+19}_{-22}\right)$ 		\\
$a_2^{\mathcal{F}_1}$	 & $-0.069\left(^{+41}_{-37}\right)$    & $ -0.068\left(^{+41}_{-30}\right)$ 		\\\hline
$a_0^{g}$	         & $0.024\left(^{+21}_{-9}\right)$      & $0.024\left(^{+12}_{-4}\right)$ 		\\
$a_1^{g}$	         & $0.05\left(^{+39}_{-72}\right)$     	& $0.05\left(^{+21}_{-41}\right)$		\\
$a_2^{g}$	         & $1.0\left(^{+0}_{-20}\right)$	& $0.9\left(^{+0}_{-18}\right)$		\\\hline\hline
\end{tabular}
\caption{Fit results using the BGL$^{(222)}$  parameterization with weak and strong unitarity bounds. 
Note that the errors are not gaussian. 
\label{tab:fit-final}}
\end{center} 
\end{table}

As mentioned in the introduction, the {\it weak} unitarity constraints of Eq.~(\ref{unitcon}) 
can be made stronger using additional information related to Heavy Quark Symmetry. In Table~\ref{tab:fit-final} we report the results of a fit that adopts the {\it strong } unitarity bounds derived in \cite{Bigi:2017jbd}.\footnote{The notation of \cite{Bigi:2017jbd} differs slightly from the present one: $a_k^{A_1}=a_k^f$, $a_k^{A_5}=a_k^{{\cal F}_1}$, $a_k^{V_4}=a_k^g$.}
  Interestingly, the results do not differ significantly from the fit with {\it weak} unitarity bounds, in contrast to analogous fits to the 2017 data only \cite{Bigi:2017jbd}.
It seems that the new and more precise data bring the fit naturally closer to the physical region,
even in the absence of strong unitarity bounds.

Another feature of the fits to the 2017  data presented in \cite{Bigi:2017njr,Bigi:2017jbd} was that the vector form factor $g(z)$ grew with $z$ (or decreased with $q^2$). This behaviour is unphysical and  led to strong deviations from the HQET expectation \cite{Bigi:2017jbd}. 
The fits in Table~\ref{tab:fit-final} do not show this pathological behaviour. 
Like in Refs.~\cite{Bigi:2017njr,Bigi:2017jbd} we also study
the inclusion of  LCSR results at $q^2=0$ in the fits, employing a 
recent  updated analysis \cite{Gubernari:2018wyi}: there seems to be excellent compatibility and $|V_{cb}|$ is basically unchanged, both with weak and strong unitarity bounds, see Table~I.

As discussed in the Introduction, at the moment we are unable to include the 
recent Babar results \cite{Dey:2019bgc} in our fit, and all previous Babar analyses 
report results only in the CLN parametrization. However, the total $B\to D^* \ell\nu$ branching fraction is essentially independent of the parametrization employed. We have checked that including the previous Babar results for the total branching fraction leaves the reference fits of Table~II almost unaffected.  

\begin{figure}[t]
	  \centering{
  \includegraphics[width=8.6cm]{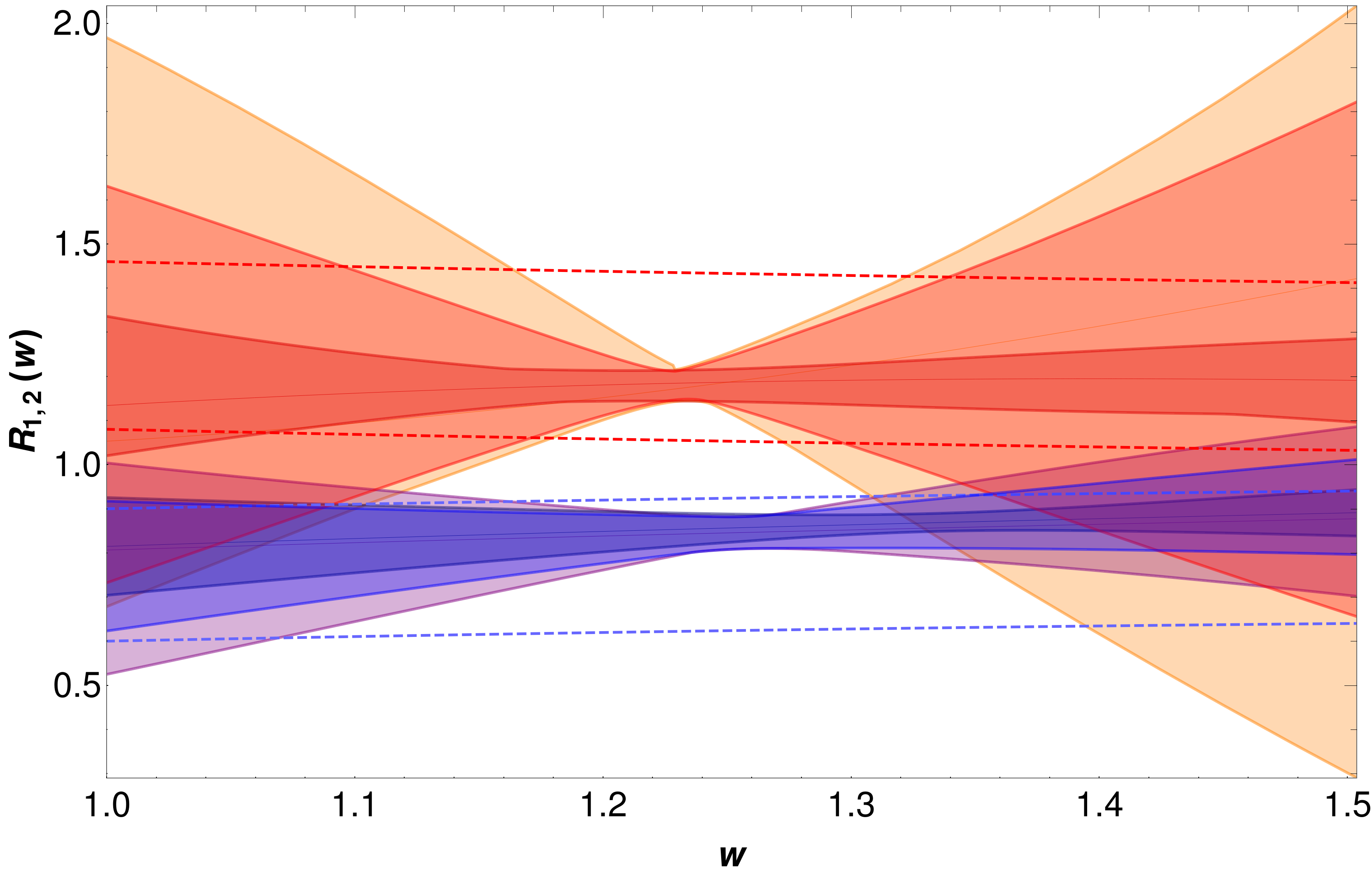} 
  \caption{Form factor ratios $R_{1,2}$ computed using the results of the fits  compared with their HQET estimate.  The three red (blue) bands show   $R_1(R_2)$ corresponding to  the fits of Table~II including  weak and strong unitarity bounds, and to  a fit with LCSR inputs and weak bounds,  in order of decreasing uncertainty. The respective HQET estimates are between the dashed lines.   }
\label{fig:hqet2}
}
\end{figure}

As already done in \cite{Bigi:2017njr,Bigi:2017jbd} we compare our results with expectations based on NLO HQET, supplemented by QCD sum rules \cite{Bernlochner:2017jka}, for which we use conservative error estimates \cite{Bigi:2017jbd}. In particular, we show results for the two ratios of form factors 
\bea
R_1(w)&=&(w+1) \,m_B m_{D^*} \frac{g(w)}{f(w)},\\
R_2(w)&=& \frac{w-r}{w-1}-\frac{{\cal F}_1(w)}{m_B (w-1) f(w)}\,.
\eea
Previous fits to the 2017 Belle data showed a marked discrepancy of $R_1$ with HQET, likely due to the unphysical behaviour of $g$ discussed above. The plot in Fig.~\ref{fig:hqet2} shows the predictions based on the fits  of Table~II (left). The uncertainties of the fit with weak unitarity constraints are as large as those in the fit to 2017 data only, but now there is everywhere good agreement with HQET. The large uncertainty in $R_1$ at low and high recoil is due to
low sensitivity to $g(z)$ in these two regions. Using the {\it strong} constraints, 
the uncertainty in those two regions decreases, and it becomes even
  smaller when using LCSR results in the fit, see Fig.~\ref{fig:hqet2}.

As mentioned above, better knowledge of the form factors in the 
small recoil region would improve significantly the determination of 
$|V_{cb}|$. In \cite{Bigi:2017njr} this was explicitly illustrated with a fit where 
we assumed that a future lattice calculation would provide the slope of ${\cal F}(w)$ 
at zero recoil.
 Here, we repeat the exercise taking inspiration  
 from the preliminary plots shown in 
 \cite{Aviles-Casco:2019vin} (although the results  are still blinded, the slope of ${\cal F}(w)$ depends only marginally on the blinding factor). We adopt $d{\cal F}/dw|_{w=1}=-1.40(7)$.
 The 5\% uncertainty appears a realistic goal for the calculations currently in progress.
 This value shows some tension with the 2018 data at small recoil, but while its inclusion in the fit increases the 
total $\chi^2$ by about 4.4, see  Table~I, it does not compromise the overall quality of the fit. At the same time $|V_{cb}|$ increases by over 1$\sigma$.
 This simple exercise does not anticipate in any way the final results of the FNAL/MILC collaboration; its only purpose is to illustrate the potential impact of lattice calculations on the fit. Inclusion of {\it strong} unitarity bounds and of LCSR does not change this picture, see Table~I.

Finally, let us comment on the binning chosen in \cite{Abdesselam:2017kjf,Abdesselam:2018nnh} for the angular variables. It is known that the single angular differential rates
have a very simple form 
that can be parametrized in terms of only 3 parameters in the cases of $\theta_l$ and $\chi$, and only 2 parameters in the case of $\theta_v$, even beyond the SM. Rather than using 10 highly correlated bins, completely  integrated over $q^2$, taking the first few moments  of $\cos \theta_{l,v}, \sin\theta_{l,v}$ or 
their analogue in $\chi$  in $q^2$ bins would  enhance the sensitivity of 
the analysis, a point emphasized also  in Ref.~\cite{Dey:2019bgc}.

\section{Semitauonic decays}
The results presented in the previous Section allow us to provide predictions for 
three quantities related to semitauonic decays: we update our predictions for
$R(D^*)$ (the ratio of semitauonic to light lepton widths) and  for the $\tau$ polarization asymmetry $P_\tau$  \cite{Bigi:2017jbd}, and we compute  the longitudinal polarization fraction of the $D^*$, $F_L^{D^*}$. There is a new form factor that enters semitauonic decays, the pseudoscalar form factor, which is unconstrained by the present experimental data and whose calculation on the lattice has not yet been completed. Here, to constrain its values we follow the third method employed in \cite{Bigi:2017jbd}: it is based on 
a kinematic relation linking it to ${\cal F}_1$ at maximum recoil and on the use of an HQET relation with conservative uncertainties at zero recoil. We obtain
\begin{align}\label{e1}
R(D^*)&=0.254^{+0.007}_{-0.006} \,,\\\label{e2}
P_\tau&=-0.476^{+0.037}_{-0.034}\,,\\ 
F_L^{D^*} &= 0.476^{+0.015}_{-0.014}\,, \label{e3}
\end{align}
where we use weak unitarity only and no LCSR input.
In comparison with \cite{Bigi:2017jbd} 
the error for $R(D^*)$ is reduced by 20\% (but remains larger than in \cite{Bernlochner:2017jka,Jaiswal:2017rve})
and the central value is about  1$\sigma$ lower.
The discrepancy of our SM prediction for $R(D^*)$ with the experimental world average 0.295(11)(8) \cite{Amhis:2016xyh} is
therefore now 2.8$\sigma$. 
On the other hand, our prediction for $P_\tau$ is almost unchanged, and of course agrees
with the experimental measurement  $P_\tau = -0.38(51)(21)$ by Belle \cite{Hirose:2017dxl}.
Our new $F_L^{D^*} $ prediction  is in good agreement with previous estimates~\cite{Fajfer:2012vx,Bhattacharya:2018kig,Murgui:2019czp} and 1.4$\sigma$ from the recent experimental measurement 
$F_L^{D^*}=0.60(8)(4)$
by the Belle collaboration \cite{Abdesselam:2019wbt}.

\section{Conclusions}

\begin{figure}[t]
	  \centering{
  \includegraphics[width=8.6cm]{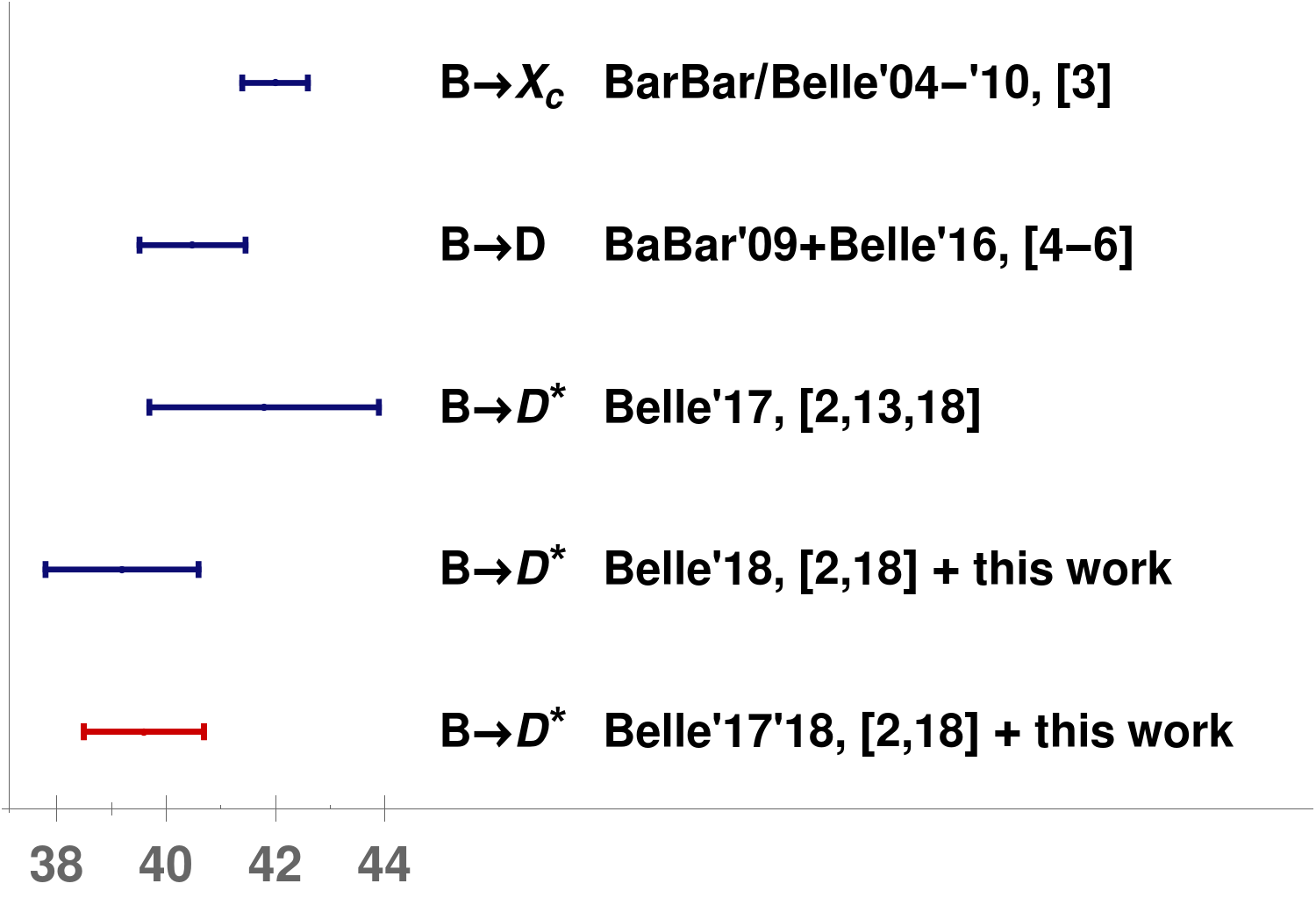} 
  \caption{Summary of $V_{cb}$ results from inclusive and exclusive decays obtained in Refs.~\cite{Gambino:2016jkc, Bigi:2016mdz, Bigi:2017njr} and this work, based on the quoted lattice QCD and experimental results. We show here the results obtained with weak unitarity constraints and no LCSR input only. \label{fig:VcbSummary} }
}
\end{figure} 

In this paper we have studied the impact of a new Belle untagged analysis of the $B\to D^* \ell\nu$ decay.
When we analyse the data of Ref.~\cite{Abdesselam:2018nnh} we obtain a value of $|V_{cb}|$ 2.1\% higher than reported there, 
with an uncertainty about 50\% larger.
Including in the fit the previous tagged analysis by Belle \cite{Abdesselam:2017kjf}
we get 
\be
|V_{cb}|= (39.6^{+1.1}_{-1.0})\times 10^{-3} \label{RDstaretc}
\ee
which still differs from the inclusive determination by about 1.9$\sigma$ and is in excellent agreement with the determination from $B\to D \ell\nu$, see the overview that we provide in Fig.~\ref{fig:VcbSummary}. We find that the inclusion of strong unitarity bounds 
and of LCSR results at maximum recoil in the fit does  not change the central value of $|V_{cb}|$, although it
helps constraining the individual form factors.  
As a byproduct of our analysis, we provide in Eqs.~(\ref{e1}--\ref{e3})
updated predictions for $R(D^*)$, $P_\tau$, and  $F_L^{D^*}$.

We also show that higher values of $|V_{cb}|$ may still be compatible with the 
available data. Indeed, preliminary results of lattice calculations suggest a 
slope of the relevant form factor ${\cal F}(w)$ at zero recoil  steeper than expected from the experimental data. We have shown that if such a high value for the slope were confirmed,
$|V_{cb}|$ extracted from a global fit to $B\to D^* \ell\nu$ data would agree with the inclusive determination.  In other words, it is lattice QCD that will decide the eventual fate of the  $|V_{cb}|$ puzzle.

\vspace{2mm}

{\bf Acknowledgements}
 We are grateful to Christoph Schwanda for useful communications concerning the Belle Collaboration results. 
 This work was supported in part by the Munich Institute for Astro- and Particle Physics (MIAPP) of the DFG cluster of excellence "Origin and Structure of the Universe". SS is supported by a DFG Forschungsstipendium under contract no. SCHA 2125/1-1, PG and MJ  by the Italian Ministry of Research (MIUR) under grant PRIN  20172LNEEZ. 
    
\bibliography{draft.bib}

\end{document}